\documentclass{ws-ijmpcs}

\usepackage{array}
\usepackage{slashed}

\DeclareMathOperator{\tr}{Tr}
\newcommand{\be}{\begin{equation}} \newcommand{\ee}{\end{equation}}
\newcommand{\ba}{\begin{eqnarray}} \newcommand{\ea}{\end{eqnarray}}
\newcommand{\bea}{\begin{eqnarray}} \newcommand{\eea}{\end{eqnarray}}
\newcommand{\bean}{\begin{eqnarray*}} \newcommand{\eean}{\end{eqnarray*}}

\newcommand{\st}{{\scriptscriptstyle T}}
\newcommand{\sL}{{\scriptscriptstyle L}}

\begin{document}

\markboth{M. G. A. Buffing and P. J. Mulders}
{Generalized Universality for TMD Distribution Functions}

%
\catchline{}{}{}{}{}
%

\title{Generalized Universality for TMD Distribution Functions}

\author{M. G. A. Buffing}

\address{Nikhef and
Department of Physics and Astronomy, VU University Amsterdam\\
De Boelelaan 1081, NL-1081 HV Amsterdam, the Netherlands\\
m.g.a.buffing@vu.nl}

\author{P. J. Mulders}

\address{Nikhef and
Department of Physics and Astronomy, VU University Amsterdam\\
De Boelelaan 1081, NL-1081 HV Amsterdam, the Netherlands\\
mulders@few.vu.nl}

\maketitle

\begin{abstract}
Azimuthal asymmetries in high-energy processes, most pronounced
showing up in combination with single or double (transverse) spin 
asymmetries, can be understood with the help of transverse
momentum dependent (TMD) parton distribution and fragmentation
functions. These appear in correlators containing expectation
values of quark and gluon operators. TMDs allow access to new
operators as compared to collinear (transverse momentum integrated)
correlators. These operators include nontrivial process dependent
Wilson lines breaking universality for TMDs. Making an angular
decomposition in the azimuthal angle, we define a set of universal
TMDs of definite rank, which appear with process dependent gluonic 
pole factors
in a way similar to the sign of T-odd parton distribution functions
in deep inelastic scattering or the Drell-Yan process.
In particular, we show that for a spin 1/2 quark target there are three pretzelocity functions.

\end{abstract}

\ccode{PACS numbers: 12.38.-t, 13.85.Ni, 13.85.Qk}

\section{Introduction}

To study the connection between partons and hadrons in high energy 
processes through parton distribution functions (PDF) and parton fragmentation 
functions (PFF), the starting points are forward matrix elements 
of parton fields, such as the quark-quark correlator
\begin{equation}
\Phi_{ij}(p|p) =
\int \frac{d^4\xi}{(2\pi)^4}\ e^{i\,p\cdot \xi}
\ \langle P\vert \overline\psi_j(0)\,\psi_i(\xi)\vert P\rangle ,
\label{phi-basic}
\end{equation}
where a summation over color indices is understood. This replaces
the correlator $\Phi \propto (\rlap{/}p + m)$ for a single incoming 
fermion. In the case of hadrons one also needs quark-quark-gluon correlators such as
\begin{equation}
\Phi^\mu_{A\,ij}(p-p_1,p_1|p) = 
\int \frac{d^4\xi\,d^4\eta}{(2\pi)^8}
\ e^{i\,(p-p_1)\cdot \xi}\ e^{i\,p_1\cdot \eta}
\ \langle P\vert \overline\psi_j(0)\,A^\mu(\eta)\,\psi_i(\xi)\vert P\rangle.
\label{quarkgluonquark}
\end{equation}
The basic idea is to factorize these hadronic (soft) parts in a full 
diagrammatic approach and parametrize them in terms of PDFs. This 
requires high energies in which case the momenta of different hadrons
obey $P{\cdot} P^\prime \propto Q^2$, where $Q^2$ is the hard scale in
the process. In that case the hadronic momenta can be treated as
light-like vectors $P$ and the hard process brings in a conjugate
light-like vector $n$ such that $P{\cdot} n = 1$, for instance
$n = P^\prime/P{\cdot} P^\prime$. One makes a Sudakov expansion of the 
parton momenta,
\begin{equation}
p = xP + p_\st + (p{\cdot} P - xM^2)n,
\end{equation}
with $x = p^+ = p{\cdot} n$. In any contraction with vectors outside the 
correlator, the component $xP$ contributes at order $Q$, the transverse
component at order $M \sim Q^0$ and the remaining component contributes
at order $M^2/Q$. This allows consecutive integration of the components 
to obtain from the
fully unintegrated correlator in Eq.~(\ref{phi-basic}) 
the TMD light-front 
(LF) correlator
\begin{equation}
\Phi_{ij}(x,p_\st;n) =
\left. \int \frac{d\,\xi{\cdot} P\,d^2\xi_\st}{(2\pi)^3}\ e^{i\,p\cdot \xi}
\ \langle P\vert \overline\psi_j(0)\,\psi_i(\xi)\vert P\rangle
\right|_{\xi{\cdot} n = 0} ,
\label{phi-TMD}
\end{equation}
the collinear light-cone (LC) correlator
\begin{equation}
\Phi_{ij}(x) =
\left. \int \frac{d\,\xi{\cdot} P}{2\pi}\ e^{i\,p\cdot \xi}
\ \langle P\vert \overline\psi_j(0)\,\psi_i(\xi)\vert P\rangle
\right|_{\xi{\cdot} n = \xi_\st = 0\ \mbox{or}\ \xi^2 = 0} ,
\label{phi-x}
\end{equation}
or the local matrix element
\begin{equation}
\Phi_{ij} =
\left. \langle P\vert \overline\psi_j(0)\,\psi_i(\xi)\vert P\rangle
\right|_{\xi = 0} .
\end{equation}
The importance of integrating at least the light-cone (minus) component 
$p^- = p{\cdot} P$ is that the expression is at equal time, 
i.e.\ time-ordering
is not relevant anymore for TMD or collinear PDFs~\cite{Diehl:1998sm}. 
For local matrix elements
one can calculate the anomalous dimensions, which show up as the Mellin 
moments of
the splitting functions that govern the scaling behavior of the 
collinear
correlator $\Phi(x)$. We note that the collinear correlator is not 
simply
an integrated TMD. The dependence on an upper limit $\Phi(x;Q^2)
= \int^{Q}d^2p_\st\ \Phi(x,p_\st)$ is found from the anomalous
dimensions (splitting functions). One has an $\alpha_s/p_\st^2$
behavior of TMDs that is calculable using collinear TMDs and which 
matches to the intrinsic 
non-perturbative $p_\st$-behavior~\cite{Collins:1984kg}.
We note that in operator product expansion language, the collinear
correlators involve operators of definite twist, while TMD correlators
involve operators of various twist.

\section{Color gauge invariance}

In order to determine the importance of a particular correlator in a 
hard process, one can do a dimensional analysis to find out when they 
contribute in an expansion in the inverse hard scale. Dominant are the 
ones with lowest canonical dimension obtained by maximizing 
contractions with $n$, for instance for quark or gluon fields the 
minimal canonical dimensions 
dim[$\overline\psi(0)\rlap{/}n\,\psi(\xi)$] = 
dim[$F^{n\alpha}(0)\,F^{n\beta}(\xi)$] = 2, while an example 
for a multi-parton combination gives 
dim[$\overline\psi(0)\rlap{/}n\,A_\st^\alpha(\eta)\,\psi(\xi)$] = 3. 
Equivalently, one can maximize the number of $P$'s in the
parametrization of $\Phi_{ij}$. Of course one immediately sees that 
any number of collinear $n{\cdot} A(\eta) = A^n(\eta)$ fields doesn't 
matter. Furthermore one must take care of color gauge invariance, for 
instance when dealing with the gluon fields and one must include 
derivatives in color gauge invariant combinations. With dimension zero 
there is $iD^n = i\partial^n + gA^n$ and
with dimension one there is $iD_\st^\alpha = i\partial_\st^\alpha
+gA_\st^\alpha$.
The color gauge-invariant expressions for quark and gluon distribution 
functions actually include gauge link operators,
\begin{equation}
U_{[0,\xi]} 
= {\cal P}\exp\left(-i\int_0^{\xi} d\zeta_\mu A^\mu(\zeta)\right),
\end{equation}
connecting the nonlocal fields,
\begin{eqnarray}
&&\Phi_{q\,ij}^{[U]}(x,p_\st;n) =
\int \frac{d\,\xi{\cdot} P\,d^2\xi_\st}{(2\pi)^3}\ e^{i\,p\cdot \xi}
\ \langle P\vert \overline\psi_j(0)\,U_{[0,\xi]}\,\psi_i(\xi)\vert P\rangle
\Biggr|_{LF} ,
\\
&&\Gamma_g^{[U,U^\prime]\,\mu\nu}(x,p_\st;n) =
{\int}\frac{d\,\xi{\cdot}P\,d^2\xi_\st}{(2\pi)^3}\ 
e^{ip\cdot\xi}
\nonumber \\ && \mbox{}\qquad\qquad\qquad \times
\tr\,\langle P{,}S|\,F^{n\mu}(0)\,
U_{[0,\xi]}^{\phantom{\prime}}\,
F^{n\nu}(\xi)\,U_{[\xi,0]}^\prime\,
|P{,}S\rangle\Biggr|_{LF} .
\end{eqnarray}
For transverse separations, the gauge links involve paths
running along the minus direction to $\pm \infty$ (dimensionally
preferred), which are closed with one or more transverse pieces 
at light-cone infinity. The two simplest
possibilities are $U^{[\pm]}$ =
$U^n_{[0,\pm\infty]}\,U^T_{[0_\st,\xi_\st]}
\,U^n_{[\pm\infty,\xi]}$,
leading to gauge link dependent quark
TMDs $\Phi_q^{[\pm]}(x,p_\st)$~\cite{Belitsky:2002sm,Bomhof:2004aw}. For gluons, the correlator involves 
color gauge-invariant traces of field-operators $F^{n\alpha}$, which 
are written in the color-triplet representation, 
requiring the inclusion of \emph{two} gauge links $U_{[0,\xi]}$ and 
$U_{[\xi,0]}^\prime$. Again the simplest possibilities are the
past- and future-pointing gauge links $U^{[\pm]}$, giving even in
the simplest case four gluon TMDs $\Gamma^{[\pm,\pm^\dagger]}(x,p_\st)$.

Using the dimensional analysis to collect the leading contributions in
an expansion in the inverse hard scale, one will need the above quark
and gluon TMDs for the description of azimuthal dependence. 
Taking the Drell-Yan (DY) process as an example, one can look at the cross 
section depending on the (small!) transverse momentum $q_\st$ of the 
produced lepton pair,
\begin{eqnarray}
\sigma(x_1,x_2,q_\st) & = & \int d^2p_{1\st}\,d^2p_{2\st}
\ \delta^2(p_{1\st}+p_{2\st}-q_\st)
\nonumber \\ && \mbox{}\hspace{1.5cm} \times 
\,\Phi^{[-]}_1(x_1,p_{1\st})
\overline\Phi^{[-^\dagger]}_2(x_2,p_{2\st})\hat\sigma(x_1,x_2,Q),
\label{XsecqT}
\end{eqnarray}
which involves a convolution of TMDs. What is
more important, it is the color flow in the process, in this case 
neutralized in the initial state, that determines the path in the gauge 
link in the TMDs, in this case past-pointing ones. In contrast in 
semi-inclusive deep inelastic scattering one finds that the relevant 
TMD is $\Phi^{[+]}$ with a 
future-pointing gauge link. In a general process one can find more 
complex gauge links including
besides Wilson line elements also Wilson loops. In particular when the
transverse momentum of more than one hadron is involved, such as 
e.g.\ in the DY case above, it may be impossible to have just a single 
TMD for a given hadron because color gets 
entangled~\cite{Rogers:2010dm,Buffing:2011mj}.

The correlators including a gauge link can be parametrized in terms of 
TMD PDFs~\cite{Bacchetta:2006tn,Mulders:1995dh} depending on $x$ and $p_\st^2$,
\begin{eqnarray}
&&\Phi^{[U]}(x,p_{\st};n) = \bigg\{
f^{[U]}_{1}(x,p_\st^2)
-f_{1T}^{\perp[U]}(x,p_\st^2)\,
\frac{\epsilon_{\st}^{p_{\st}S_{\st}}}{M}
+g^{[U]}_{1s}(x,p_\st)\gamma_{5}
\nonumber \\&&\mbox{}\qquad
+h^{[U]}_{1T}(x,p_\st^2)\,\gamma_5\,\slashed{S}_{\st}
+h_{1s}^{\perp [U]}(x,p_\st)\,\frac{\gamma_5\,\slashed{p}_{\st}}{M}
+ih_{1}^{\perp [U]}(x,p_\st^2)\,\frac{\slashed{p}_{\st}}{M}
\bigg\}\frac{\slashed{P}}{2},
\label{e:par}
\end{eqnarray}
with the spin vector parametrized as 
$S^\mu = S_{\sL}P^\mu + S^\mu_{\st} + M^2\,S_{\sL}n^\mu$ 
and shorthand notations for $g^{[U]}_{1s}$ and $h_{1s}^{\perp [U]}$,
\begin{equation}
g^{[U]}_{1s}(x,p_\st)=S_{\sL} g^{[U]}_{1L}(x,p_{\st}^2)
-\frac{p_{\st}\cdot S_{\st}}{M}g^{[U]}_{1T}(x,p_{\st}^2).
\end{equation}
For quarks, these include not only the 
functions that survive upon $p_\st$-integration, $f_1^q(x) = q(x)$, 
$g_1^q(x) = \Delta q(x)$ and $h_1^q(x) = \delta q(x)$, which are the 
well-known collinear spin-spin densities 
(involving quark and
nucleon spin) but also momentum-spin densities such as the Sivers 
function
$f_{1T}^{\perp q}(x,p_\st^2)$ (unpolarized quarks in a transversely 
polarized nucleon) 
and spin-spin-momentum densities such as $g_{1T}(x,p_\st^2)$ 
(longitudinally polarized quarks in a transversely polarized nucleon). 

The parametrization for gluons,
following the naming convention in Ref.~\refcite{Meissner:2007rx}, 
is given by
\begin{eqnarray}
&&2x\,\Gamma^{\mu\nu [U]}(x{,}p_\st) = 
-g_T^{\mu\nu}\,f_1^{g [U]}(x{,}p_\st^2)
+g_T^{\mu\nu}\frac{\epsilon_T^{p_TS_T}}{M}
\,f_{1T}^{\perp g[U]}(x{,}p_\st^2)
\nonumber\\&&\mbox{}\qquad
+i\epsilon_T^{\mu\nu}\;g_{1s}^{g [U]}(x{,}p_\st)
+\bigg(\frac{p_T^\mu p_T^\nu}{M^2}\,
{-}\,g_T^{\mu\nu}\frac{p_\st^2}{2M^2}\bigg)\;h_1^{\perp g [U]}(x{,}p_\st^2)
\nonumber\\ &&\mbox{}\qquad
-\frac{\epsilon_T^{p_T\{\mu}p_T^{\nu\}}}{2M^2}\;
h_{1s}^{\perp g [U]}(x{,}p_\st)
-\frac{\epsilon_T^{p_T\{\mu}S_T^{\nu\}}
{+}\epsilon_T^{S_T\{\mu}p_T^{\nu\}}}{4M}\;
h_{1T}^{g[U]}(x{,}p_\st^2).
\label{GluonCorr}
\end{eqnarray}

\section{Transverse moments}

In many cases, it is convenient to construct moments of TMDs in the 
same way as one considers moments of collinear functions. 
For $\Phi(x)$ in Eq.~(\ref{phi-x}) one constructs moments
\begin{eqnarray}
x^{N}\Phi(x) & = & 
\left. \int \frac{d\,\xi{\cdot} P}{2\pi}\ e^{i\,p{\cdot} \xi}
\ \langle P\vert \overline\psi(0)\,(i\partial^n)^{N}
\,U^n_{[0,\xi]}\,\psi(\xi)\vert P\rangle
\right|_{LC} 
\nonumber \\ & = & 
\left. \int \frac{d\,\xi{\cdot} P}{2\pi}\ e^{i\,p{\cdot} \xi}
\ \langle P\vert \overline\psi(0)\,U^n_{[0,\xi]}\,(iD^n)^{N}
\,\psi(\xi)\vert P\rangle
\right|_{LC}.
\end{eqnarray}
Integrating over $x$ one finds the connection of the Mellin moments 
of PDFs
with local matrix elements with specific anomalous dimensions,
which via an inverse Mellin transform define the splitting functions.
Similarly one can consider transverse moment weighting starting with 
the light-front TMD in Eq.~(\ref{phi-TMD}),
\begin{eqnarray}
p_\st^\alpha\,\Phi^{[\pm]}(x,p_\st;n) & = & 
\int \frac{d\,\xi{\cdot} P\,d^2\xi_\st}{(2\pi)^3}
\ e^{i\,p{\cdot} \xi}
\nonumber\\ && \mbox{}\hspace{-0.1cm} \times 
\langle P\vert \overline\psi(0)\,U^n_{[0,\pm\infty]}
\,U^T_{[0_\st,\xi_\st]}
\,iD_\st^\alpha(\pm\infty)
\,U^n_{[\pm\infty,\xi]}\psi(\xi)\vert P\rangle
\Biggr|_{LF} .
\end{eqnarray}
Integrating over $p_\st$ gives the lowest transverse moment, which 
appears in the $q_\st$-weighted result of Eq.~(\ref{XsecqT}). This moment 
involves 
twist-3 (or higher) collinear multi-parton correlators, in particular 
the quark-quark-gluon correlator
\begin{eqnarray}
\Phi^{n\alpha}_{F}(x-x_1,x_1|x) & = &
\int \frac{d\,\xi{\cdot} P\,d\,\eta{\cdot} P}{(2\pi)^2}
\ e^{i\,(p-p_1){\cdot} \xi}\ 
\nonumber\\ && \mbox{}\hspace{0.5cm} \times 
e^{i\,p_1{\cdot} \eta}
\ \langle P\vert \overline\psi(0)\,U^n_{[0,\eta]}\,F^{n\alpha}(\eta)
\,U^n_{[\eta,\xi]}\,\psi(\xi)\vert P\rangle\Biggr|_{LC}.
\end{eqnarray}
In terms of this correlator and the similarly defined correlator 
$\Phi_D^\alpha(x-x_1,x_1|x)$ one finds 
\begin{equation}
\int d^2p_\st\ p_\st^\alpha\,\Phi^{[U]}(x,p_\st)
= \tilde\Phi_\partial^\alpha(x) + C_G^{[U]}\,\pi\,\Phi_G^\alpha(x),
\end{equation}
with
\begin{eqnarray}
&&\tilde\Phi_\partial^\alpha(x)=\Phi_D^\alpha(x) - \Phi_A^\alpha(x)
\nonumber \\ &&\mbox{}\hspace{1.5cm} 
= \int dx_1\,\Phi_D^\alpha(x-x_1,x_1 | x)
-\int dx_1\,\text{PV}\frac{1}{x_1}\,\Phi_F^{n\alpha}(x-x_1,x_1 | x),
\nonumber
\\&&
\Phi_G^\alpha(x) = \Phi_F^{n\alpha}(x,0 | x).
\nonumber
\end{eqnarray}
The latter is referred to as a gluonic pole or ETQS-matrix 
element~\cite{Efremov:1984ip,Qiu:1991pp}. 
They are multiplied with gluonic pole factors $C_G^{[U]}$ 
(e.g.\ $C_G^{[\pm]} = \pm 1$), that tell us that new functions are
involved with characteristic process dependent 
behavior~\cite{Bacchetta:2005rm,Bomhof:2006ra}. This behavior
is for the single transverse moments also coupled to the behavior
under time-reversal. 
While $\tilde\Phi_\partial^\alpha$ is T-even, $\Phi_G^\alpha$ is T-odd.
Since time-reversal is a good symmetry of QCD, the appearance of T-even
or T-odd functions in the parametrization of the correlators is linked
to specific observables with this same character. In particular single
spin asymmetries are T-odd observables.

The weighting with transverse momenta can also be analyzed by studying 
the parametrization in PDFs. 
For single $p_{\st}$-weighting, only PDFs with one prefactor of 
$p_{\st}$ in the parametrization in Eq.~(\ref{e:par}) survive. 
The $\widetilde\Phi_{\partial}^{\alpha}(x)$ matrix element receives contributions from 
T-even PDFs, while the $\Phi_{G}^{\alpha}(x)$ matrix element receives contributions from T-odd PDFs, see Ref.~\refcite{Boer:2003cm} for 
a detailed study of this. For the single weighted results, thus, the behavior under
time-reversal can be used to identify the 
process dependent parts and we find
\begin{equation}
f_{1T}^{\perp(1)[U]}(x)=C_G^{[U]} f_{1T}^{\perp(1)}(x),
\label{signchange}
\end{equation}
where transverse weighting for PDFs is defined as
\be
f_{\ldots}^{(n)}(x) 
= \int d^2 p_{\st}\left(\frac{-p_\st^2}{2M^2}\right)^n\,f_{\ldots}(x,p_\st^2)
\label{transversemoments}
\ee
for weighting with $n$ transverse moments.
The fact that depending on the process $\Phi^{[\pm]}$ are the 
correlators to be used, leads to the sign change~\cite{Collins:2002kn} for the
T-odd functions in such processes or to more complex factors if
more complex gauge links are involved~\cite{Bomhof:2006dp}. The importance of
Eq.~(\ref{signchange}) is the appearance of 
a universal function with 
calculable process (link) dependent numbers rather than 
many process dependent functions that are somehow related. 
For gluon TMDs, there are
already for single weighting two functions $\Gamma_G^{(f/d)}$ and
hence two different gluonic pole factors $C_G^{[U](f/d)}$, because
there are two ways to construct color singlets from the (in that
case) three gluon fields that are involved using the $f_{abc}$ or
$d_{abc}$ structure constants. The appearance of two different
gluon Sivers functions was pointed out in Ref.~\refcite{Bomhof:2007xt}.

The situation with universality for fragmentation functions~\cite{Collins:2004nx} is easier because
the gluonic pole matrix elements vanish in that 
case~\cite{Meissner:2008yf,Gamberg:2008yt,Gamberg:2010gp}. 
Nevertheless, there 
exist T-odd fragmentation functions, but their QCD operator structure 
is T-even. These T-odd functions then appear in the parametrization
of $\tilde\Phi_\partial^\alpha$.
Hence, there is no process dependence from gluonic pole factors.

The use of transverse moments in the description of azimuthal
asymmetries via transverse momentum weighting of the cross section
can be extended to higher moments involving higher 
harmonics such as $\cos(2\varphi)$. Also here process dependence
may come in from double gluonic pole matrix elements
$\Phi_{GG}^{\alpha\beta}$, which are twist four operators. 
This affects studies that involve the quark 
TMD $h_{1T}^{\perp q}(x,p_\st)$ (pretzelocity distribution)
in Eq.~(\ref{e:par}) or the gluon Boer-Mulders function 
$h_1^{\perp g}(x,p_\st)$ (linearly polarized gluons in 
unpolarized targets) in Eq.~(\ref{GluonCorr}).
For instance, for quarks one finds for the simplest
gauge links,
\be
h_{1T}^{\perp(2)[\pm]}(x)=h_{1T}^{\perp(2)(A)}(x)
+ h_{1T}^{\perp(2)(B1)}(x),
\label{threepretzel}
\ee
where the functions $h_{1T}^{\perp(2)(A)}(x)$, 
$h_{1T}^{\perp(2)(B1)}(x)$ are 
universal. Actually the latter function corresponds to
a correlator $\Phi_{GG}$, involving a color structure
${\rm Tr}_c \left[FF\psi\overline\psi\right]$. For more complex 
gauge links one actually needs a third (universal) pretzelocity
second transverse moment because there is another 
possible color structure~\cite{Buffing:2012sz}.

\section{TMDs of definite rank}

An interesting possibility to obtain universal TMDs is to
start with a parametrization that involves the symmetric 
traceless tensors $p_\st^{\alpha_1\ldots\alpha_m}$
of rank $m$, such as 
\begin{equation}
p_\st^\alpha, \quad p_\st^{\alpha\beta} = p_\st^\alpha\,p_\st^{\beta} 
- \frac{1}{2}p_\st^2\,g_\st^{\alpha\beta} \ldots \, .
\end{equation} 
Depending on the rank different correlators come in, involving
operator combinations of gluons, covariant derivatives and $A$-fields.
Minimizing the twist we have
\bea
\Phi^{[U]}(x,p_\st) &\ =\ &
\Phi(x,p_\st^2) 
+ \pi C_{G}^{[U]}\,\frac{p_{\st i}}{M}
\,\Phi_{G}^{i}(x,p_\st^2)
+ \pi^2 C_{GG,c}^{[U]}\,\frac{p_{\st ij}}{M^2}
\,\Phi_{GG,c}^{ij}(x,p_\st^2)
+\ldots
\nonumber \\ &\quad +&
\frac{p_{\st i}}{M}\,\widetilde\Phi_\partial^{i}(x,p_\st^2)
+ \pi C_{G}^{[U]}\,\frac{p_{\st ij}}{M^2}
\,\widetilde\Phi_{\{\partial G\}}^{\,ij}(x,p_\st^2)
+ \ldots
\nonumber \\ &\quad +&
\frac{p_{\st ij}}{M^2}
\,\widetilde\Phi_{\partial\partial}^{ij}(x,p_\st^2)
+ \ldots \, ,
\label{TMDstructure}
\eea
with a summation over the color structures $c$.
The reproduction of the transverse moments provides the proper
identification of universal TMD functions,
\bea
&&\Phi(x,p_\st^2) = 
\bigg\{
f_{1}(x,p_\st^2)
+S_\sL\,g_{1}(x,p_\st^2)\gamma_{5}
+h_{1}(x,p_\st^2)\gamma_5\,\slashed{S}_{\st}
\bigg\}\frac{\slashed{P}}{2},
\label{rank0}
\\
&&\frac{p_{\st i}}{M}\,\widetilde\Phi_\partial^{i}(x,p_\st^2) =
\bigg\{
h_{1L}^{\perp}(x,p_\st^2)\,S_\sL\frac{\gamma_5\,\slashed{p}_{\st}}{M}
-g_{1T}(x,p_\st^2)\,\frac{p_\st{\cdot}S_\st}{M}\gamma_{5}
\bigg\}\frac{\slashed{P}}{2},
\\ &&
\frac{p_{\st i}}{M}\,\Phi_{G}^{i}(x,p_\st^2)
= \frac{1}{\pi}\bigg\{
-f_{1T}^{\perp}(x,p_\st^2)\,
\frac{\epsilon_{\st}^{\rho\sigma}p_{\st\rho}S_{\st\sigma}}{M}
+ih_{1}^{\perp}(x,p_\st^2)\,\frac{\slashed{p}_{\st}}{M}
\bigg\}\frac{\slashed{P}}{2},
\\
&&
\frac{p_{\st ij}}{M^2}\widetilde\Phi_{\partial\partial}^{ij}(x,p_\st^2)
= h_{1T}^{\perp (A)}(x,p_\st^2)
\,\frac{p_{\st ij}S_\st^i\,\gamma_5\gamma_{\st}^j}{M^2}
\,\frac{\slashed{P}}{2},
\\ &&
\frac{p_{\st ij}}{M^2}\Phi_{GG,1}^{ij}(x,p_\st^2)
= \frac{1}{\pi^2}h_{1T}^{\perp (B1)}(x,p_\st^2)
\,\frac{p_{\st ij}S_\st^i\,\gamma_5\gamma_{\st}^j}{M^2}
\,\frac{\slashed{P}}{2},
\\ &&
\frac{p_{\st ij}}{M^2}\Phi_{GG,2}^{ij}(x,p_\st^2)
= \frac{1}{\pi^2}h_{1T}^{\perp (B2)}(x,p_\st^2)
\,\frac{p_{\st ij}S_\st^i\,\gamma_5\gamma_{\st}^j}{M^2}
\,\frac{\slashed{P}}{2},
\\ &&
\frac{p_{\st ij}}{M^2}\widetilde\Phi_{\{\partial G\}}^{ij}(x,p_\st^2)
= 0. \label{PhipartialG}
\eea
We note that the rank zero functions in Eq.~(\ref{rank0}) depend on
$x$ and $p_\st^2$ and involve traces,
to be precise
$g_1(x,p_\st^2) = g_{1L}^{[U]}(x,p_\st^2)$ and
$h_1(x,p_\st^2) = h_{1T}^{[U]}(x,p_\st^2) 
- (p_\st^2/2M^2)\,h_{1T}^{\perp [U]}(x,p_\st^2)$. 
As remarked before, for the pretzelocity there are three
universal functions with in general
\be 
h_{1T}^{\perp [U]}(x,p_\st^2)
= h_{1T}^{\perp (A)}(x,p_\st^2)
+ C_{GG,1}^{[U]}\,h_{1T}^{\perp (B1)}(x,p_\st^2)
+ C_{GG,2}^{[U]}\,h_{1T}^{\perp (B2)}(x,p_\st^2).
\ee
For the simplest gauge links we have $C_{GG,1}^{[\pm]} = 1$ and
$C_{GG,2}^{[\pm]} = 0$, which shows e.g.\ that 
$h_{1T}^{\perp [{\rm SIDIS}]}(x,p_\st^2)
=h_{1T}^{\perp [{\rm DY}]}(x,p_\st^2)$, but that for other
processes (with more complicated gauge links) other
combinations of the three possible pretzelocity functions occur.
For a spin 1/2 target the above set of TMDs is complete. There are
no higher rank functions. For a spin 1 target~\cite{Bacchetta:2000jk,Hoodbhoy:1988am} and for gluons, there
are higher rank functions~\cite{Buffing:2012sz,BMM2}.
For the fragmentation correlator there is for rank 2 only a single 
(T-even) pretzelocity function 
$H_{1T}^\perp(z,k_\st^2)$ appearing in the parametrization of the 
correlator $\Delta_{\partial\partial}^{\alpha\beta}(x,p_\st^2)$.
The rank of the various correlators is shown in 
Table~\ref{t:spinhalfcol} 
with the results for nucleon TMD PDFs summarized in 
Tables~\ref{t:unpol} (unpolarized) and
\ref{t:spinhalfpol} (polarized) and those for nucleon TMD PFFs
in Tables~\ref{t:spinhalffragunp} and~\ref{t:spinhalffragpol}. 

\begin{table}[!tb]
\tbl{The correlators of definite rank in the full TMD correlator, ordered in columns according to the number of gluonic poles ($G$) and ordered in rows according to the number of contributing partial derivatives ($\partial$ = $D-A$). The rank of these operators is equal to the sum of these numbers. Their twist is equal to the rank + 2.}
{\begin{tabular}{|m{26mm}|m{26mm}|m{26mm}|m{26mm}|m{26mm}|}
\hline
\multicolumn{4}{|c|}{GLUONIC POLE RANK} \\ \hline
\qquad\quad 0 & \qquad\quad 1 & \qquad\quad 2 & \qquad\quad 3 
\\ \hline
$\Phi(x,p_\st^2)$
&$\pi C_{G}^{[U]}\,\Phi_{G}$
&$\pi^2 C_{GG,c}^{[U]}\,\Phi_{GG,c}$
&$\pi^3 C_{GGG,c}^{[U]}\,\Phi_{GGG,c}$
\\[2pt]
\hline
$\widetilde\Phi_\partial$
&$\pi C_{G}^{[U]}\,\widetilde\Phi_{\{\partial G\}}$
&$\pi^2 C_{GG,c}^{[U]}\,\widetilde\Phi_{\{\partial GG\},c}$
& \ldots
\\[2pt]
\hline
$\widetilde\Phi_{\partial\partial}$
&$\pi C_{G}^{[U]}\,\widetilde\Phi_{\{\partial\partial G\}}$
& \ldots & \ldots
\\[2pt]
\hline
$\widetilde\Phi_{\partial\partial\partial}$
& \ldots & \ldots & \ldots
\\[2pt] 
\hline
\end{tabular}
\label{t:spinhalfcol}}
\end{table}

\begin{table}[!tb]
\centering
\begin{minipage}{0.45\linewidth}
\tbl{TMD PDFs for an unpolarized or spin 0 target assigned to the quark correlators as given in Table~\ref{t:spinhalfcol}.}
{\begin{tabular}{|m{15mm}|m{15mm}|m{15mm}|}
\hline
\multicolumn{3}{|c|}{PDFs FOR SPIN 0 HADRONS} \\ \hline
$f_1$
&$h_{1}^{\perp}$
&
\\[2pt]
\hline
&
\\[2pt]
\cline{1-2}
\\[2pt]
\cline{1-1}
\end{tabular}
\label{t:unpol}}
\end{minipage}
\hspace{8mm}
\begin{minipage}{0.45\linewidth}
\tbl{Assignments of TMD PDFs for a polarized spin 1/2 target.\vspace{3.5mm}}
{\begin{tabular}{|m{13mm}|m{12mm}|m{21mm}|}
\hline
\multicolumn{3}{|c|}{PDFs FOR SPIN 1/2 HADRONS} \\ \hline
$g_{1}$, $h_{1}$
&$f_{1T}^{\perp}$
&$h_{1T}^{\perp(B1)}$, $h_{1T}^{\perp(B2)}$
\\[2pt]
\hline
$g_{1T}$, $h_{1L}^{\perp}$
&
\\[2pt]
\cline{1-2}
$h_{1T}^{\perp(A)}$
\\[1pt]
\cline{1-1}
\end{tabular}
\label{t:spinhalfpol}}
\end{minipage}
\\[6mm]
\begin{minipage}{0.45\linewidth}
\tbl{Assignments of TMD PFFs for unpolarized or spin 0 hadrons.}
{\begin{tabular}{|m{15mm}|m{15mm}|m{15mm}|}
\hline
\multicolumn{3}{|c|}{PFFs FOR SPIN 0 HADRONS} \\ \hline
$D_1$
&
&
\\[2pt]
\hline
$H_{1}^{\perp}$
&
\\[2pt]
\cline{1-2}
\\[1pt]
\cline{1-1}
\end{tabular}
\label{t:spinhalffragunp}}
\end{minipage}
\hspace{8mm}
\begin{minipage}{0.45\linewidth}
\tbl{Assignments of TMD PFFs for polarized spin 1/2 hadrons.}
{\begin{tabular}{|m{22mm}|m{11.5mm}|m{11.5mm}|}
\hline
\multicolumn{3}{|c|}{PFFs FOR SPIN 1/2 HADRONS} \\ \hline
$G_{1}$, $H_{1}$
&
&
\\[2pt]
\hline
$G_{1T}$, $H_{1L}^{\perp}$, $D_{1T}^{\perp}$
&
\\[2pt]
\cline{1-2}
$H_{1T}^{\perp}$
\\[1pt]
\cline{1-1}
\end{tabular}
\label{t:spinhalffragpol}}
\end{minipage}
\end{table}

\section{Conclusions}
We have introduced quark TMD correlators of definite rank in Eq.~(\ref{TMDstructure}). In this new decomposition, we have made an expansion of the quark correlator into irreducible tensors multiplying correlators containing operator combinations of gluons, covariant derivatives and $A$-fields, the latter in the combination
$i\partial = iD - gA$. In the decomposition gluonic pole factors
contain the gauge link dependence, which are calculated from
the transverse moments. These factors also give the process
dependence, which is determined by the gauge link structure.
The correlators of definite rank, in turn are parameterized in terms
of the universal TMD PDFs depending on $x$ and $p_\st^2$, such as given 
by Eqs.~(\ref{rank0})-(\ref{PhipartialG}). The process dependence for
a particular TMD PDF is in the same gluonic pole factors that appear 
in the expansion in Eq.~(\ref{TMDstructure}).

An analysis for a quark spin 1/2 target shows that the process 
dependence is not strictly confined to the T-odd functions, such as the 
Sivers or the Boer-Mulders functions. In fact, we have shown the 
existence of three T-even pretzelocity functions. For fragmentation the 
TMD PFFs are already universal since gluonic pole matrix elements
vanish for fragmentation correlators.
Future work will be focused on the study of universality for higher
spin targets and gluon TMD PDFs~\cite{Buffing:2012sz,BMM2}. While for a spin 1/2 target one 
has at most rank two TMDs, one has for higher spins and gluon TMDs
also higher rank functions, while also the color and gauge link
structure is richer.

\section*{Acknowledgments}
Part of this work is in collaboration with Asmita Mukherjee. We acknowledge discussions with Jianwei Qiu.
This research is part of the research program of the Foundation
for Fundamental Research of Matter (FOM),  
which is financially supported by the Netherlands Organisation
for Scientific Research (NWO).
MGAB also acknowledges
support of the FP7 EU-programme HadronPhysics3 (contract
no 283286).

\end{document}